\documentclass[aps,prl,twocolumn,letterpaper,superscriptaddress,showpacs]{revtex4}
\usepackage{graphicx}
\usepackage{CJK}

\begin{document}
\begin{CJK*}{UTF8}{bsmi}
\title{
Can disorder alone destroy the $e_g^{\prime}$ hole pockets of Na$_{x}$CoO$_{2}$?
\\
A Wannier function based first-principles method for disordered systems
}
\author{Tom Berlijn}
\affiliation{Physics Department, State University of New York, Stony Brook,
New York 11790, USA}
\affiliation{Condensed Matter Physics and Materials Science Department,
Brookhaven National Laboratory, Upton, New York 11973, USA}
\author{Dmitri Volja}
\altaffiliation{current address: Department of Materials Science and Engineering, MIT, Cambridge, MA 02139, USA.}
\affiliation{Physics Department, State University of New York, Stony Brook,
New York 11790, USA}
\affiliation{Condensed Matter Physics and Materials Science Department,
Brookhaven National Laboratory, Upton, New York 11973, USA}
\author{Wei Ku(%
顧威)
}
\affiliation{Condensed Matter Physics and Materials Science Department,
Brookhaven National Laboratory, Upton, New York 11973, USA}
\affiliation{Physics Department, State University of New York, Stony Brook,
New York 11790, USA}

\date{\today}

\begin{abstract}
We investigate from first principles the proposed destruction of the controversial $e_g^{\prime}$ pockets in the Fermi surface
 of Na$_x$CoO$_2$ due to Na disorder, by calculating its $k$-dependent configuration-averaged spectral function $\langle A(k,\omega)\rangle$.
To this end, a Wannier function-based method is developed that treats the effects of disorder beyond the mean field.
Remarkable spectral broadenings of order $\sim$1eV are found for the oxygen orbitals, possibly explaining their absence in the experiments.
In contradiction with the current claim, however, the $e_g^{\prime}$ pockets remain almost perfectly coherent.
The developed method is expected to also generate exciting opportunities in the study of the countless functional materials that owe their important electronic properties to disordered dopants.

\end{abstract}


\maketitle
\end{CJK*}


Like most transition metal oxides, sodium cobaltates obtain their important properties via the introduction of dopants (Na) between the layered structures of oxygen and transition metal atoms.
Around $x=0.3$, Na$_x$CoO$_2$ develops unconventional superconductivity under hydration~\cite{Takada}, with evidence of a nodal order                    parameter~\cite{Zheng}.
From $x<0.5$ to $x>0.5$ it changes from a paramagnetic metal to a Curie-Weiss metal,
while it is a charge ordered insulator at $x=0.5$~\cite{Foo}.
At high doping, the combination of high thermopower and high conductivity is observed~\cite{Terasaki}, together with
A-type antiferromagnetism~\cite{Bayrakci}.
In addition, various Na orderings have been observed throughout the entire phase diagram~\cite{Zandbergen}.
This rich variety of behaviors has thus attracted intensive research activity.

Nevertheless, even the most basic starting point for an understanding is still under serious debate, namely the low-energy electronic structure near the chemical potential that controls most of the above mentioned remarkable properties.
Early density functional (DFT) calculations~\cite{Singh1} within the local density approximation (LDA) predicted the existence of a central $a_g$ hole pocket, surrounded by 6 $e_g^{\prime}$ hole pockets.
Angle resolved photoemission spectroscopy (ARPES) experiments~\cite{Hasan20061568,Yang}  measured the central $a_g$ pocket, but found the $e_g^{\prime}$ bands to be below the Fermi surface.
Shubnikov-de Haas measurements~\cite{Balicas} observed two pockets, but the assignment to $e_g^{\prime}$ was concluded incompatible with the specific heat data~\cite{Ueland}.
On the other hand, the presence of the second type of pocket was reconfirmed by Compton scattering~\cite{Laverock} and assigned to $e_g^{\prime}$.

Despite the controversial status on the experimental front, various theoretical efforts have been made to investigate the alleged absence of the $e_g^{\prime}$ pocket.
Surface effects were suggested to suppress the $e_g^{\prime}$ pocket~\cite{Pillay2} under hydroxyl contamination.
The other usual suspect of many-body correlation was investigated within dynamical mean-field theory (DMFT) by several groups.
However, the results were inconclusive as the $e_g^{\prime}$ pockets were found to either grow or shrink depending sensitively on the crystal field splitting~\cite{Lechermann}.
Some researchers~\cite{Marianetti2} argued that the $e_g^{\prime}$ pockets should not exist according to the specific heat data, while others~\cite{Pillay} concluded from an extensive study of the crystal field that the $e_g^{\prime}$ pockets cannot be removed via local correlation.

Recently, an intriguing alternative resolution was proposed~\cite{Singh2}.
It was argued that the random positioning of the Na intercalants alone can introduce strong disorder effects that mask the $e_g^{\prime}$ pockets from the ARPES experiments.
This physically plausible picture, if proven, would not only enable a new resolution to reconcile the various theoretical and experimental observations, but would also introduce important novel physics missing in current considerations.  
Unfortunately, verification of this intriguing proposal presents a great challenge to the current capability of the first-principles theories.
To unambiguously resolve the complete disorder effects with a good $k$-space resolution, the disorder-induced self-energy cannot be assumed local \textit{a priori}. Especially with the potential localization\cite{Singh2} near the pockets, the spatial fluctuations need to be taken into account, beyond the current state-of-the-art mean-field treatments~\cite{Stocks,Lechermann2,Marianetti1}.

In this Letter, we examine the proposal of disorder-induced destruction of the $e_g^{\prime}$ pockets in Na$_x$CoO$_2$, using $x=0.3$ as a representative case, by developing a first-principles Wannier function-based method for the evaluation of electronic structure of disordered materials.
Remarkable spectral broadenings ($\sim$1eV) of the oxygen bands are found that provide a natural explanation for the missing oxygen bands in ARPES studies.
However, in contradiction with the current claim~\cite{Singh2}, we only find a negligible influence of disorder on the $e_g^{\prime}$ pockets, incapable of masking them from the ARPES experiments.
The inclusion of nonlocal disorder-induced self-energy in our method is expected to generate exciting opportunities in the studies of countless modern functional materials, including doped transition metal oxides, dilute magnetic semiconductors, and intercalated graphites, to name but a few.


The electronic band structure of a disordered system is determined by the configuration-averaged spectral function:
$\langle A(k,\omega)\rangle$=$\sum_{c}P_{c}A_{c}(k,\omega)$ of momentum $k$ and energy $\omega$,
 in which configuration $c$ is weighted by its probability $P_{c}$.
By treating the disordered configurations within the supercell approximation,
the spectral functions $A_{c}(k,\omega )$ can be calculated directly
from the supercell eigenvalues and eigenvectors, by applying the unfolding method of~\cite{unfolding}.
While conceptually clean, the approach is in practice too computationally expensive for most applications
 due to the large size of the supercells required to incorporate the characteristic length scale of localization and impurity scattering.
The goal of this work, therefore, is to drastically reduce the computational cost
by developing an effective low-energy disordered Hamiltonian using first-principles Wannier functions.

The Hamiltonian of an arbitrary configuration of $N$ impurities, positioned at $(x_{1},...,x_{N})$, can be exactly rewritten as
\begin{eqnarray}\label{eq:eq1}
H^{(x_{1},...,x_{N})}=H^{0}+\sum_{i=1}^{N}\Delta^{(x_{i})}+\sum_{i> j=1}^{N}\Delta^{(x_{i},x_{j})}+...
\end{eqnarray}
 where $H^{0}$ denotes the Hamiltonian of the system with no impurities,
 $\Delta^{(x_{i})}$=$H^{(x_{i})}$-$H^{0}$, denotes the linear influence of the impurity at $x_{i}$
 and $\Delta^{(x_{i},x_{j})}$=$H^{(x_{i},x_{j})}$-$\Delta^{(x_{i})}$-$\Delta^{(x_{j})}$-$H^{0}$ denotes the two-body correction
 of a pair of impurities at ($x_{i}$,$x_{j}$), etc.
Naturally, low-order Green's functions like the spectral functions here are not too sensitive to higher order corrections
 and the computation can be simplified accordingly.
In practice we found that it is already highly accurate to keep only the linear influence of the impurities, as demonstrated below.
It is important to note that a cutoff here by no means implies a truncation in the electronic multiple scattering processes among the impurities.
Also, while Eq.~(\ref{eq:eq1}) resembles the well established cluster expansion method~\cite{Hinuma} of the formation energy of large-sized impurity clusters, it encapsulates the influence of disorder in the entire low-energy effective Hamiltonian.

The construction of the effective Hamiltonian explicitly consists of three steps.
First, DFT calculations are performed for the undoped normal cell, and for each type of impurity in a large supercell containing one impurity.
Second, the influence of each impurity (located at $x_m$ in the first normal cell) is extracted, $\Delta^{(m)}$=$H^{(m)}$-$H^{0}$, in the basis of Wannier functions.
Similar to the induced forces in the typical frozen phonon calculations, the partition of the influence of the impurity from its superimages is necessary~\cite{SUP}.
Third, the effective low-energy Hamiltonian corresponding to each disorder configuration with $N$ impurities is then assembled:
\begin{eqnarray}\label{eq:eq2}
&&\langle r'n'| H_{\rm{eff}}^{\{(r_{1},m_{1}),...,(r_{N},m_{N}) \} } | r''n''\rangle =
 \\
&&\langle r'-r'',n' |H^{0}|0n'' \rangle
+\sum_{i=1}^{N}
\langle r'-r_i,n' |\Delta^{(m_i)}|r''-r_i,n'' \rangle \;  \nonumber
\end{eqnarray}
where $r$ and $n$ denote the lattice vector and the orbital index of the Wannier functions, and $r_i$ and $m_i$ denote the lattice vector and the type of impurity $i$, located at $x_i=r_i+x_{m_i}$. The notation here also implies possible addition of impurity orbitals and removal of orbitals due to vacancies. Obviously, a similar procedure can be performed if higher order contributions are desired.

Note that only when the supercell and normal cell Hamiltonians are represented on the same Wannier basis can one legally add and subtract them.
Therefore, the Wannier functions should be constructed from a large energy window such that the Hilbert space is as complete as possible.
Furthermore, the projected Wannier function method~\cite{Ku, Anisimov} naturally ensures maximum consistency of the normal cell and supercell~\cite{SUP}, and is thus more suitable than the maximally localized Wannier function method~\cite{Vanderbilt} which risks defining the gauge differently in the supercell in favor of better localization near the impurity.

\begin{figure}
\includegraphics[width=1.0\columnwidth,clip=true]{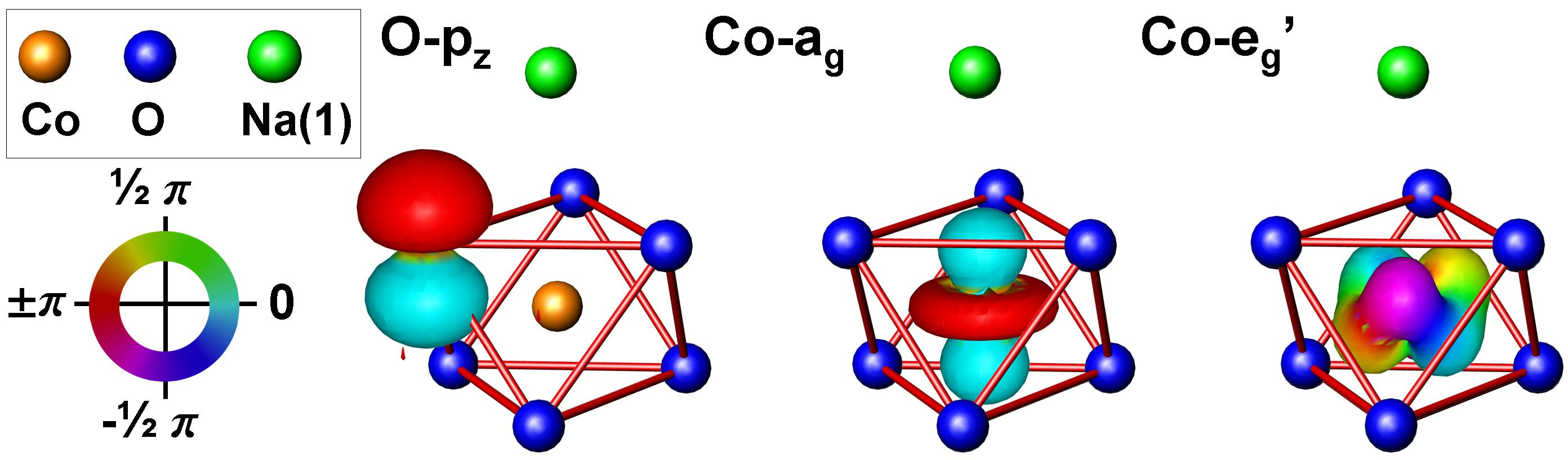}
\caption{\label{fig:fig1}
(color online)
Isosurface plot $|\langle x|rn\rangle|=0.09(\rm{bohr})^{-3/2}$ of selected Wannier functions, colored according to the phase of the complex functions.
}\end{figure}

For the case of Na$_x$CoO$_2$, we consider both possible locations of Na between the oxygen layers discussed previously in the literature~\cite{Singh2}: above Co [Na(1)] or above a hole in the Co sheet [Na(2)].  The low-energy Hilbert space is taken within [-7,3] eV consisting of symmetry-respecting complex Wannier orbitals of Co-$d$ and O-$p$ characters (22/normal cell)~\cite{SUP}, as illustrated in Fig.~\ref{fig:fig1}.  The effects of impurities are extracted from three DFT calculations~\cite{SUP}: the undoped Co$_2$O$_4$ in the normal cell and Na(1)Co$_8$O$_{16}$ and Na(2)Co$_8$O$_{16}$ in $2\times 2\times 1$ supercells corresponding to $x=\frac{1}{8}$.  Since effects of lattice relaxation on the $e_g^{\prime}$ pockets are found to be negligibly small in comparison~\cite{SUP}, we proceed with the experimentally averaged structure of Na$_{0.3}$CoO$_2$ in the following analysis.

\begin{figure}
\includegraphics[width=1.0\columnwidth,clip=true]{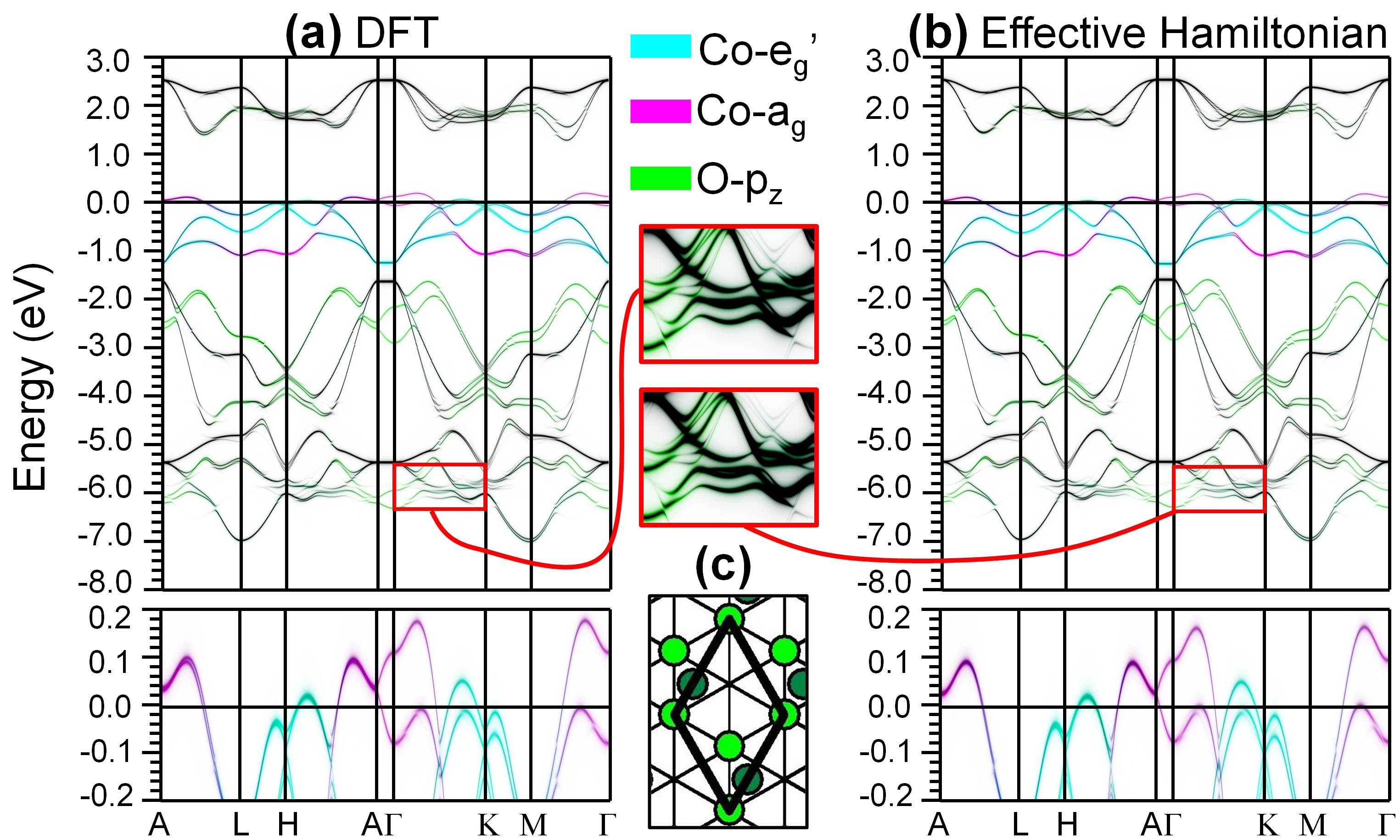}
\caption{\label{fig:fig2}
(color online)
Spectral function of Na(2)Na(1)$_{2}$Co$_{6}$O$_{12}$ obtained from (a) the full DFT calculation, on a basis of 2019 LAPW's
(b) a single diagonalization of the effective Hamiltonian, on a basis of 66 Wannier functions. (c) Na positions in the supercell, in relation to triangular Co sheet.  Light(dark) circles denote Na above(below) the Co sheet.
}\end{figure}

The quality of our effective Hamiltonian can be verified by benchmarking the spectral function of a test system, against the standard DFT.
As a highly nontrivial test case we take the periodic Na(2)Na(1)$_{2}$Co$_{6}$O$_{12}$(cf. Fig.\ref{fig:fig2}(c)), which requires a strongly ``incommensurate'' extrapolation in the partitioning (from $2\times 2\times 1$ to $\sqrt{3}\times\sqrt{3}\times 1$ cell) and the linearity (from $x=\frac{1}{8}$ to $x=\frac{1}{6}+\frac{2}{6}$) of the influence of impurities.
As shown in Fig.~\ref{fig:fig2}, our effective Hamiltonian manages to reproduce the spectral function of the full DFT calculation with high accuracy (in particular the details around the gap opening), but with only a negligible fraction of the computational effort.
(The full DFT calculation involved $\sim$20 self-consistent cycles on a basis of 2019 linear augmented plane waves, while our effective Hamiltonian requires only a single diagonalization on a basis of 66 Wannier functions.) Additional benchmarks exploring the potential limitations of our approximations for large extrapolations are given in Ref.~\cite{SUP}.

\begin{figure}
\includegraphics[width=1.0\columnwidth,clip=true]{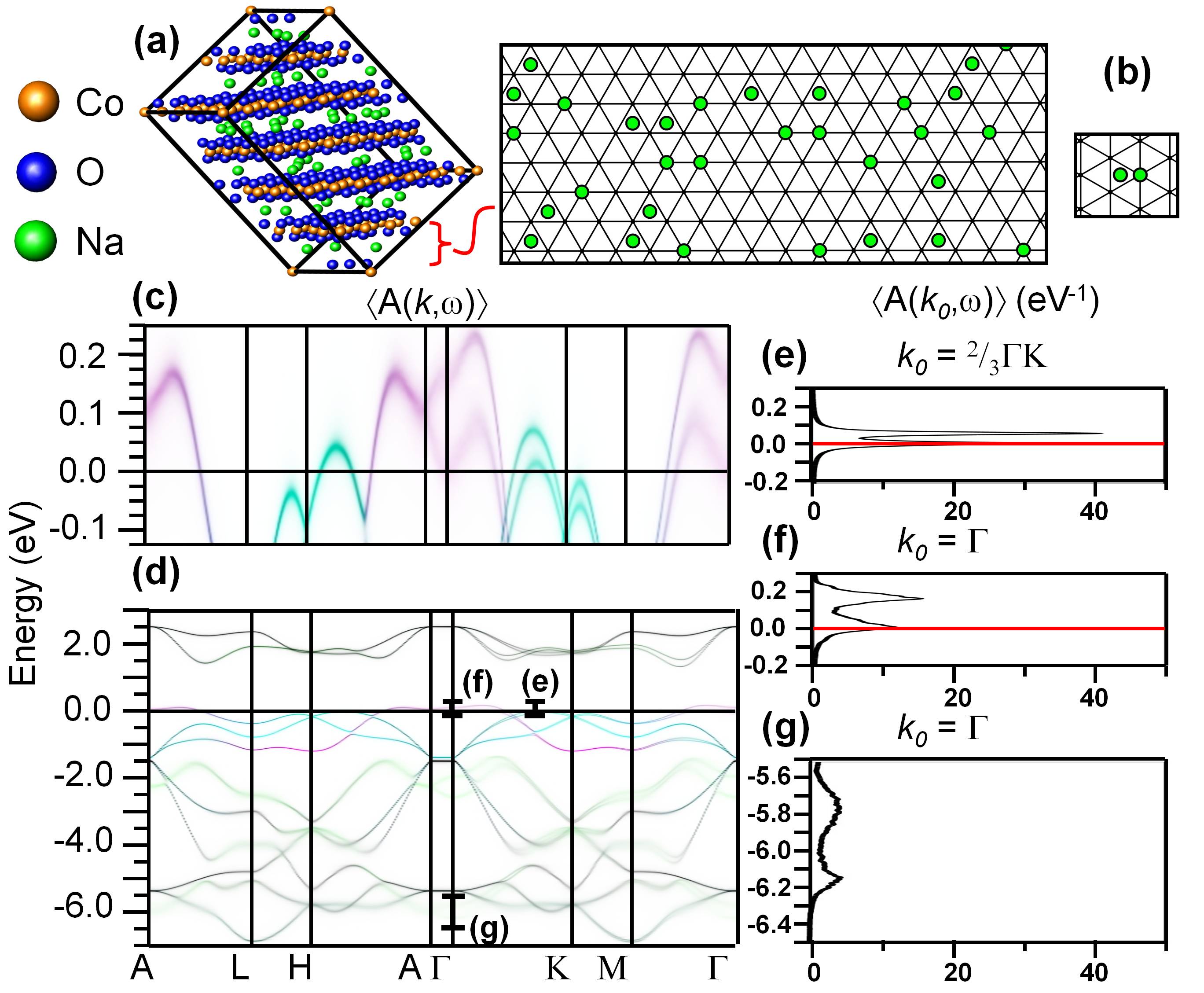}
\caption{\label{fig:fig3}
(color online)
Configuration-averaged spectral function of Na$_{0.3}$CoO$_2$, showing the $e_g^{\prime}$ states to be the least influenced by disorder.
(a) One of the 50 large-sized supercells used for configurational average.
(b) A high energy configuration with Na(1) too close to Na(2).
$\langle A(k,\omega)\rangle$ (c) around the Fermi surface (d) in the full low-energy Hilbert space, where the bars correspond to the energy distribution curves
(e) at $k_{0}$=2$\Gamma$K/3,
(f) at $k_{0}$=$\Gamma$ ,
(g) at $k_{0}$=$\Gamma$.
The Wannier orbital character is labeled according to the color scheme of Fig.~\ref{fig:fig2}.
}\end{figure}

Having an accurate and efficient method to assemble the effective Hamiltonian of any configuration, we proceed to evaluate the configuration-averaged spectral functions for the case of disordered Na$_{0.3}$CoO$_2$.
Following the considerations laid out in Ref.~\cite{Singh2}, all the random configurations of Na are assumed comparable in probability, except the high energy case containing two nearest neighboring Na atoms located at Na(1) and Na(2) sites [Fig.~\ref{fig:fig3}b], which is disregarded due to its low probability.
Figure~\ref{fig:fig3} shows the resulting spectral function converged with respect to the number of configurations (50) and their average size ($\sim$80 normal cells corresponding to $80\times22=1760$ Wannier functions).
Note that it is necessary to consider supercells [e.g.,Fig.~\ref{fig:fig3}(a)] of different sizes, orientations, and shapes in order to remove the effects of artificial zone boundaries of the supercell.


A remarkable broadening of oxygen bands can be observed in Figs.~\ref{fig:fig3}(d)(g), indicating a short lifetime and mean free path of the quasiparticles due to strong scattering against the disordered Na atoms.
This is understandable considering that the Na atoms are located in the oxygen cages defined by the two oxygen layers, and thus have the largest impact on the oxygen orbitals.
Interestingly, this huge spectral broadening and low intensity might explain why some of the oxygen bands are not observed in the ARPES measurements~\cite{Yang, Qian1}, not only in Na$_x$CoO$_2$, but also in most doped layered transition metal oxides, where the dopants are introduced between the oxygen layers.

In great contrast, much weaker effects of disorder are found on the Co-$d$ orbitals [cf. Figs.~\ref{fig:fig3}(c)(e)(f)].  Specifically, the $a_g$ orbital picks up some $k$-{\itshape dependent} broadening near the $\Gamma$ point, while the $e_g^{\prime}$ orbitals are almost unaffected by the disorder.  This somewhat surprising result probably reflects the strong screening of oxygen that shields the cobalt valence orbitals from the influence of the disordered Na, and the metallic nature of the doped electrons that spread throughout the whole system.
Clearly, the localization and broadening of the $e_g^{\prime}$ and $a_g$ bands are not strong enough to shift the $e_g^{\prime}$ pockets below the Fermi level.
Unlike the $a_g$ orbital that points directly toward the doped Na atom (cf.: Fig.~\ref{fig:fig1}), the $e_g^{\prime}$ orbitals neither point toward the Na atom nor the most affected O-$p_z$ orbitals near the Na, making them the least sensitive to the presence of Na intercalants.
This could also explain the negligible effects of lattice relaxation around Na atoms on the $e_g^{\prime}$ pockets~\cite{SUP}.
Obviously, the disorder alone does \textit{not} destroy the six $e_g^{\prime}$ Fermi pockets of Na$_{0.3}$CoO$_2$,
in contradiction with the current claim~\cite{Singh2}.

On the other hand, the Na impurities do introduce an important physical effect on the Co-$d$ shell, namely on the crystal field splitting (the relative on-site-energy) of the $e_g^{\prime}$ and $a_g$ orbitals.  Indeed, evaluated from $H^{0}$ and $\Delta^{Na(1)}$, the crystal field splitting is found to change from 27 meV to -38 meV for Co atoms right below or above the Na intercalants, consistent with the trend estimated previously from the density of states~\cite{Marianetti1}.
Considering the tendency of strong orbital polarization of the many-body exchange interactions as demonstrated from the previous DMFT studies~\cite{Lechermann,Marianetti1}, the combination of disorder and strong exchange interactions is very likely to give stronger scattering for the $e_g^{\prime}$ orbitals.  Exactly whether this would lead to a resolution of the highly controversial status of the $e_g^{\prime}$ pockets of Na$_{x}$CoO$_2$, remains an interesting challenge to the theorists.

Interestingly, our results also demonstrate a significant nonlocality of the disorder-induced self-energy.  Indeed, a strongly k-dependent spectral broadening can be clearly observed in Fig.~\ref{fig:fig3} that correlates well with the inverse of the band velocity rather than the energy.  For example, at 25 meV the almost purely $a_g$ bands have large ($\sim$200meV) spectral broadening near the $\Gamma$ point, but negligible width at $k\sim\frac{1}{2}$A$\Gamma$.  Such a strong $k$-dependence of the spectral width reflects the intrinsic nonlocality of the self-energy, and highlights the advancement of our method over standard mean-field theories in which the self-energy is assumed local.

\begin{figure}
\includegraphics[width=1.0\columnwidth,clip=true]{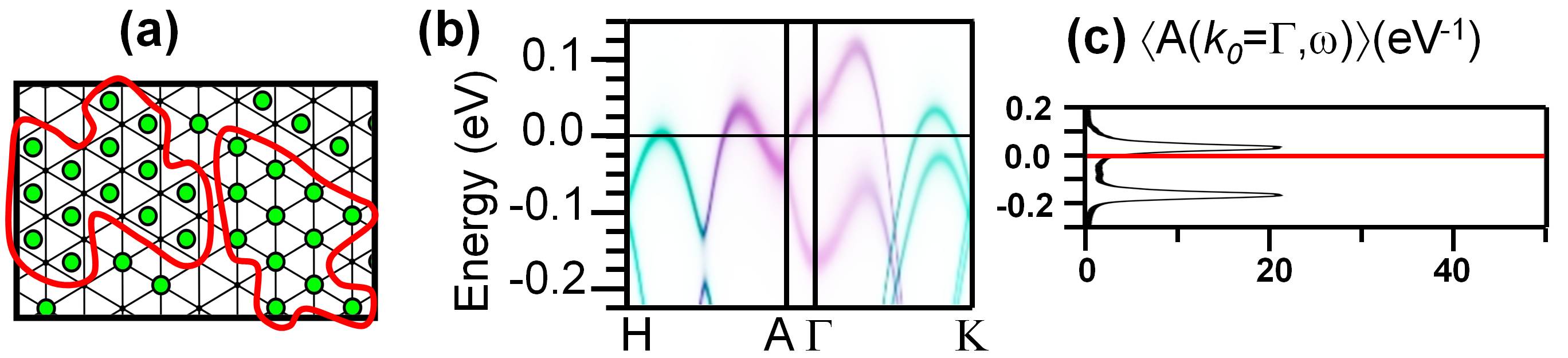}
\caption{\label{fig:fig4}
(color online)
Configuration-averaged spectral function of Na$_{0.7}$CoO$_{2}$,
 showing short-range order suppressing the spectral broadening of $a_g$.
(a) Small islands of homogeneous Na(1) and Na(2).
(b) $\langle A(k,\omega)\rangle$ around the Fermi-surface.
(c) energy distribution curve at $k_{0}$=$\Gamma$.
The Wannier orbital character is labeled according to the color scheme of Fig.~\ref{fig:fig2}.
}\end{figure}

Amazingly, related to the nonlocal self-energy, we also found significant effects of short-range ordering of Na impurities known to be important for the Na$_x$CoO$_2$~\cite{Zandbergen,Hinuma}.  As demonstrated in Fig.~\ref{fig:fig4}(a) for Na$_{0.7}$CoO$_2$, the exclusion of nearest neighbor Na positioning introduces automatically a strong short-range ordering of the Na impurities, due to lack of available locations at high doping (see also \cite{SUP}).  In turn, the resulting spectral function of the $a_g$ orbitals, for example, demonstrates stronger coherence and longer lifetime, as can be seen by comparing Figs.~\ref{fig:fig4}(b) and ~\ref{fig:fig4}(c) with Figs.~\ref{fig:fig3}(c) and ~\ref{fig:fig3}(f). (The $e_g^{\prime}$ orbitals near the pockets remain perfectly coherent also at this doping level.)  This result, while physically intuitive, is actually quite nontrivial, since in the mean-field theories the degree of disorder at 0.7 doping level should be exactly the same as that at 0.3 doping level.  

In conclusion, we have investigated from first principles the proposed destruction of the controversial $e_g^{\prime}$ pockets in the Fermi surface of Na$_x$CoO$_2$ due to Na disorder.
To this end, a Wannier function-based method is developed that incorporates the spatial distributions of impurities beyond the mean field.
The new method is benchmarked against the full DFT calculation and shown to be efficient and highly accurate.
Remarkable $k$-dependent broadenings of the spectral function are found in the oxygen orbitals due to their vicinity to the Na intercalants.
However, the effects of disorder are found to be negligible on the $e_g^{\prime}$ orbitals.  Thus, the disorder alone does not destroy the $e_g^{\prime}$ pockets,
in contradiction with the current claim~\cite{Singh2}.
Interestingly, against the mean-field perspective, enhanced coherence is found at higher doping where short-range order grows stronger.
Our new method is expected to find a wide range of applications in the studies of the countless functional materials that owe their important electronic properties to disordered dopants.

This work was supported by the U.S. Department of Energy, Basic Energy Sciences, Materials Sciences and Engineering Division, and DOE-CMSN.

\bibliography{refr3v1}
\end{document}